\documentclass{article}

\usepackage[final,nonatbib]{neurips_2022}

\usepackage[square,sort,comma,numbers]{natbib}
\usepackage[utf8]{inputenc} 
\usepackage[T1]{fontenc}    
\usepackage{hyperref}       
\usepackage{url}            
\usepackage{booktabs}       
\usepackage{amsfonts}       
\usepackage{nicefrac}       
\usepackage{microtype}      
\usepackage{xcolor}         

\setcitestyle{square}
\usepackage{graphicx}
\usepackage{cleveref}
\usepackage{wrapfig}

\title{Does Medical Imaging learn different Convolution Filters?}

\author{%
  Paul Gavrikov$^{1}$ \qquad Janis Keuper$^{1,2}$\\
  $^{1}$IMLA, Offenburg University, Germany \\
  $^{2}$Fraunhofer ITWM, Kaiserslautern, Germany\\
  \texttt{\{paul.gavrikov, janis.keuper\}@hs-offenburg.de} \\
}

\begin{document}

\maketitle
\begin{abstract}
Recent work has investigated the distributions of learned convolution filters through a large-scale study containing hundreds of heterogeneous image models. 
Surprisingly, on average, the distributions only show minor drifts in comparisons of various studied dimensions including the learned task, image domain, or dataset. However, among the studied image domains, medical imaging models appeared to show significant outliers through ``spikey'' distributions, and, therefore, learn clusters of highly specific filters different from other domains. 
Following this observation, we study the collected medical imaging models in more detail. We show that instead of fundamental differences, the outliers are due to specific processing in some architectures. Quite the contrary, for standardized architectures, we find that models trained on medical data do not significantly differ in their filter distributions from similar architectures trained on data from other domains. Our conclusions reinforce previous hypotheses stating that pre-training of imaging models can be done with any kind of diverse image data. 
\end{abstract}

\section{Introduction}
Deep learning has accelerated the progress in many computer vision problems, including various applications in medical imaging \cite{7404017, Suzuki2017Sep, Ronneberger2015Nov,7444155}. Most deployed architectures are based on convolutional neural networks (CNNs) which learn large amounts of convolution filters to transform inputs into meaningful feature spaces that can be further processed.
Since the same CNN architectures can learn many computer vision tasks and work on diverse image domains, the question naturally arises whether there are shared similarities in the learned convolution filters.

Previously, we introduced a large-scale dataset (\texttt{CNN Filter DB}) \cite{Gavrikov_2022_CVPR} that consists of convolution filters with a kernel size of $3\times 3$ extracted from hundreds of publicly available pre-trained CNNs trained for a multitude of tasks. On that data, we performed a singular-value decomposition (SVD) comparison of sets of filters aggregated by various dimensions of meta-data. Therefore, sets of normalized filters were transformed to a common SVD-obtained basis, and the shift between resulting distributions of the coefficients ($c_{ij}$) was measured based on a variant of the Kullback-Leibler Divergence (KL). Generally, individual models and model families (e.g. \texttt{ResNet}) learn unique distributions of filters, but on average, distributions segregated by the learned task, image domain, or dataset do not significantly differ. Most observed shifts were due to different levels of ``degenerated'' (i.e. sparse or highly repetitive) filters caused by the over-parameterization of networks relative to the dataset.
However, the most salient outlier in shifts included medical imaging models, as this domain showed distinct ``spikey'' coefficient distributions. In this paper, we set our focus to study these models in detail and understand the possible causes of the shifts. The models can be summarized in the following:
\texttt{CompNet} \cite{dey2018compnet} contains three customized architectures trained for brain segmentation on the \texttt{OASIS} (MRI) dataset \cite{Marcus2007Sep};
\texttt{LungMask} \cite{hofmanninger2020automatic} contains three \texttt{UNet} trained for lung segmentation on the \texttt{LTRC} and some proprietary CT datasets;
\texttt{TorchXRayVision}  \cite{Cohen2022xrv} contains seven \texttt{DensetNet-121} and one \texttt{ResNet50} classifier trained to detect various pathologies from various datasets. Additionally, the framework includes one \texttt{ResNet101} auto-encoder trained on a large aggregated dataset consisting of the previously mentioned datasets;
Lastly, one individual \texttt{UNet} \cite{buda2019association} is included that is also trained to perform brain segmentation on a public \texttt{Kaggle} MRI dataset \cite{Shih2019Jan}.

\section{Analysis}

\begin{figure}
    \centering
    \includegraphics[width=\linewidth]{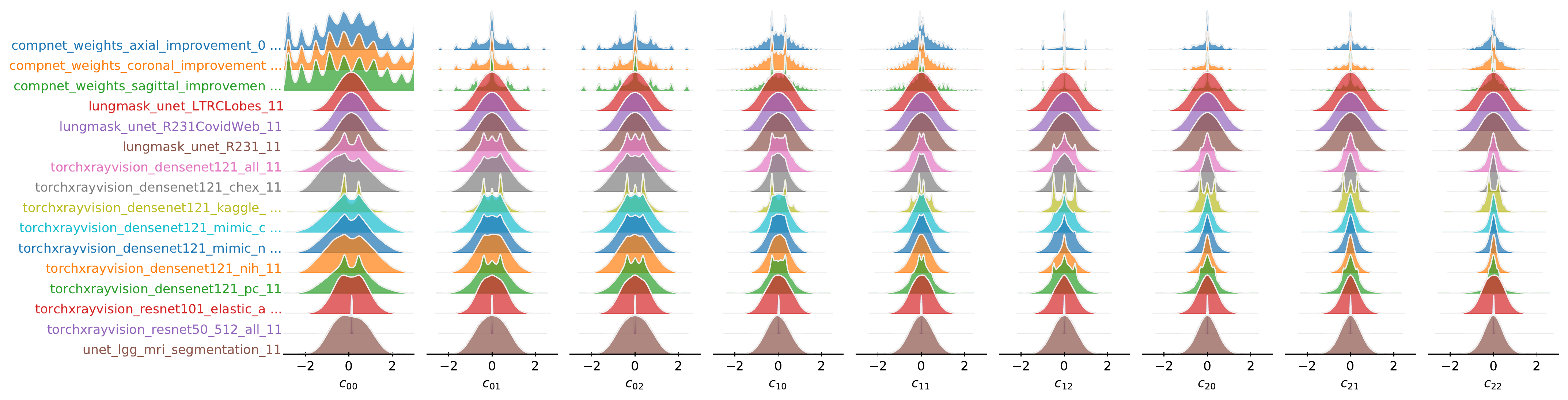}
    \caption{KDEs of the coefficient distributions along every principal component for all medical imaging models from \texttt{CNN Filter DB}.}
    \label{fig:kdes_models}
\end{figure}

In this section, we apply the same SVD-decomposition as defined in \cite{Gavrikov_2022_CVPR} to visualize the coefficient distributions in kernel density estimation (KDE) plots. We collect all filters for each medical model and collectively visualize the resulting coefficients (\Cref{fig:kdes_models}).
First, we observe that the salient \underline{``spikey'' distributions are only seen in models of \texttt{CompNet}} on all axes, while all other models form less salient distributions. In the original analysis, this model family was over-represented due to their relatively large number of convolutions.

The \texttt{CompNet} architecture can be split into multiple encoder-decoder models: source inputs are fed into an encoder that feeds into two parallel decoder branches. One of those constructs a binary brain segmentation mask, and the other computes a complementary output for the non-brain part of the image. Finally, the outputs are aggregated and further processed by an auto-encoder that learns a latent-space representation. 
\begin{wrapfigure}{r}{0.3\linewidth}
    \centering
    \includegraphics[width=\linewidth]{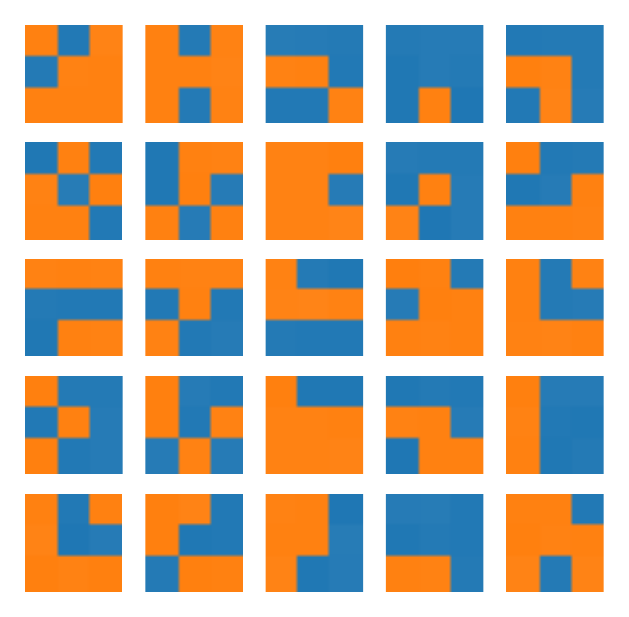}
    \caption{The first filters from a layer that learns only binary-like filters found in \texttt{CompNet}.}
    \label{fig:binary_filters}
\end{wrapfigure}
Upon closer analysis, we observe that some layers learn filters with binary weights (\Cref{fig:binary_filters}) which explains the coefficient spikes in the KDEs, while other layers learn a less conspicuous set of filters. We assume that the binary filters are learned by the final auto-encoder as indicated by the presence in later convolution stages, yet, we are unable to prove our hypothesis due to the highly entangled implementation of the source model. Still, it seems worthwhile to understand the cause of binary filters as it is possible to learn them from a significantly lower dimensional and discrete search space which should accelerate learning and require fewer data samples. Interestingly, we find fewer such binary filter layers in the model trained on the axial reconstruction plane.

Next, complementary to the findings in \cite{Gavrikov_2022_CVPR}, we observe that models based on the same architecture (e.g. \texttt{DenseNet121}) learn quite similar distributions when compared to each other but differ substantially when compared to other architectures: all \texttt{CompNets} learn similar coefficient distributions although trained on different reconstruction planes; Albeit trained in different regimes, all \texttt{LungMask} models also form similar distributions to the standalone \texttt{UNet}.  Still, some shifts in the distributions can be seen, in particular in the \texttt{DenseNet121} models from \texttt{TorchXRayVision} which were trained on different datasets. The model trained on the \texttt{Kaggle RSNA Pneumonia Challenge} dataset seems to overrepresent specific coefficients (and therefore filters clusters) by a large amount, while the models trained on other datasets such as \texttt{MIMIC-CXR} \cite{data_mimic}, \texttt{NIH} \cite{data_nih,data_nih_google}, and \texttt{CheXpert} \cite{data_chexpert} tend to learn coefficient distributions that are smoother and more similar to the distribution of the model that was trained on all datasets. The most salient coefficient distribution is seen in the \texttt{ResNet50} which was trained on $512\times 512$ px images instead of $256\times 256$ px. Spikes dominate the KDEs at 0 for all axes, which indicates a highly sparse model. Indeed, by applying a structured pruning on the kernels, we find that only approx. 1\% of filters (N=12900) is non-sparse. Due to a lack of access to the datasets, we cannot verify the integrity of this model. However, if this model's performance is on-par with the other models, this opens the question of whether a) smaller and therefore easier-to-train models would not be more suitable for this problem and perhaps brain segmentation in general and b) higher resolution images may be more suitable as they may contain more salient features for classification than their down-sampled counter-parts.


\begin{figure}
    \centering
    \includegraphics[width=0.8\linewidth]{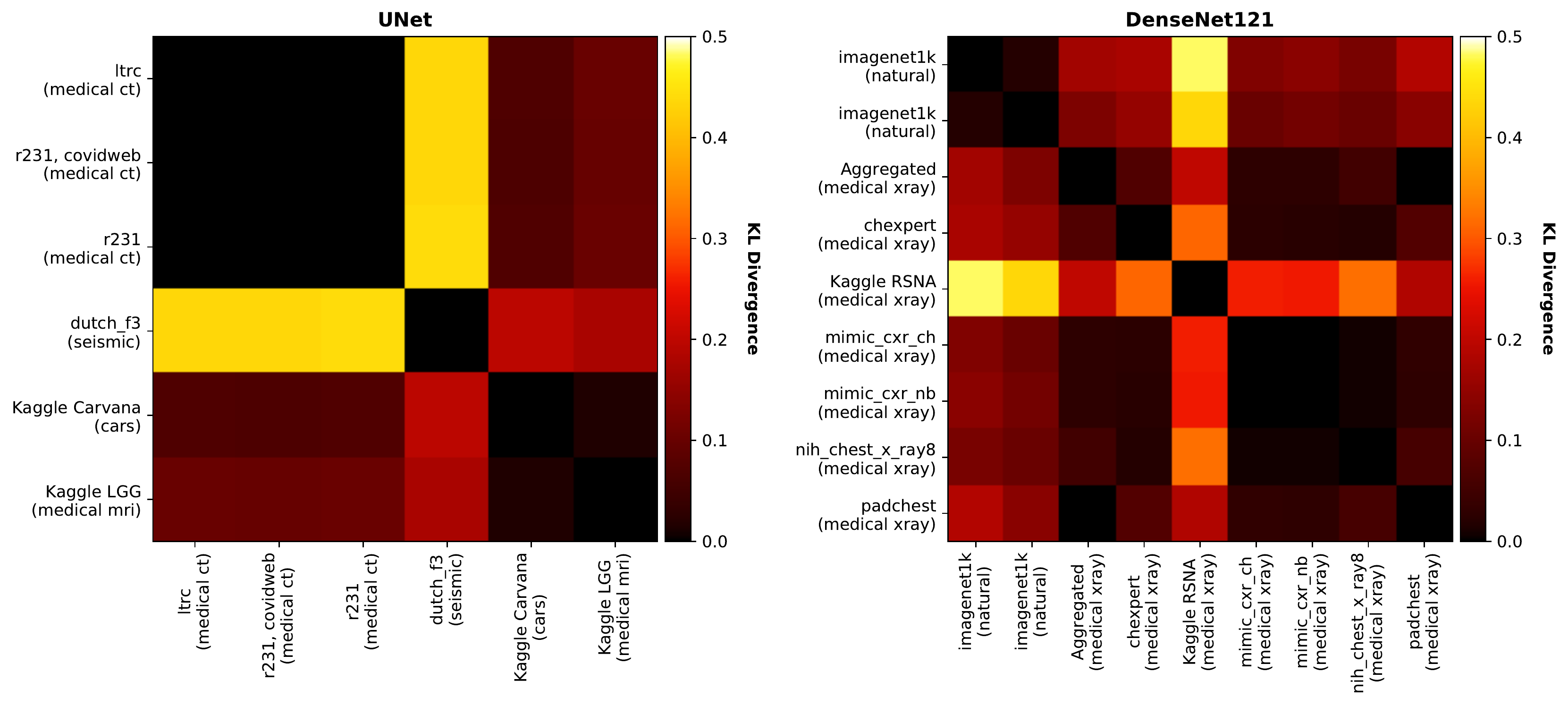}\hspace{2em}
    \caption{Heatmaps showing the pair-wise distribution shift for \texttt{UNet} (left) and \texttt{DenseNet} (right) trained on different datasets. Low values/dark colors denote low shifts.}
    \label{fig:kl_combined}
\end{figure}

Finally, we compare the medical \texttt{UNet} segmentation and \texttt{DenseNet121} classification models to similar models from \texttt{CNN Filter DB} trained on other image domains (cars, seismic data, and natural images) to understand whether medical models learn different convolution filters. We compute the KL in a pair-wise manner between all model combinations and display the results in the form of heatmaps (\Cref{fig:kl_combined}). We observe that \texttt{UNet} learns strikingly different representations on seismic data, but only minor differences when trained to segment cars. \texttt{DenseNet121} also learns slightly different representation when trained on \texttt{ImageNet} \cite{deng2009imagenet}. Yet, it is worth noting that this is a significantly larger dataset with 1.4 M samples and 1000 categories that occupy more of the model capacity. 
Finally, the most noticeable shift is seen in the model trained on the \texttt{Kaggle} dataset, although it originates from the same image domain. 

In all cases, it appears as if the shifts between filters of medical domains and others are insignificant and may even vanish when compared to more models. 

\section{Conclusion and Future Work}

The previously observed outliers in distributions \cite{Gavrikov_2022_CVPR} are directly caused by specific branches in architectures, such as \cite{dey2018compnet}. After all, it turns out that medical imaging models do not learn fundamentally different filter distributions than models of other image domains. In turn, this means that pre-training with diverse image data originating from arbitrary domains should accelerate the training of medical models (and others). The remaining shifts to other domains may be explained by different levels of over-parameterization and bear potential for future work which may study the filter quality through metrics proposed in \cite{Gavrikov_2022_CVPR, Gavrikov_2022_CVPRb}. For comparison, a larger sample size of models should be used.

Additionally, we have observed that some layers learn highly specific clusters of filters. Others, seem to develop extreme levels of sparsity. In those cases, it would be reasonable to reduce the search space during training, and in turn, accelerate training, and/or reduce the amount of necessary training data. Given the scarcity of medical data and its acquisition cost, it seems worthwhile to explore this direction in future work.

\section*{Potential negative societal impact}
Our paper is limited to the continued analysis of learned models, so we do not believe our findings cause any negative societal impact. However, on contrary, we hope that they can accelerate the learning of medical imaging models which would result in reduced CO$_{2}$ emissions during training and increase AI democratization due to the reduced amount of necessary training data.

\begin{ack}
Funded by the Ministry for Science, Research and Arts, Baden-Wuerttemberg, Germany, Grant 32-7545.20/45/1 (Q-AMeLiA).
\end{ack}
\medskip
{
\small
\bibliographystyle{IEEEtran}
\bibliography{main}
}

\end{document}